\documentclass[%
 reprint,
 superscriptaddress,
 showpacs,
 amsmath,amssymb,
 aps,
 pra,
 floatfix,
 longbibliography
]{revtex4-1}

\usepackage{graphicx}% Include figure files
\usepackage{dcolumn}% Align table columns on decimal point
\usepackage{bm}% bold math
\usepackage{braket}
\usepackage[colorlinks=true,allcolors=blue]{hyperref}

\begin{document}

%\preprint{APS/123-QED}

\title{Controllable spin-Hall and related effects of light in an atomic medium via coupling fields}

\affiliation{Department of Physics, Zhejiang University, Hangzhou 310027, China}
\affiliation{Institute for Quantum Science and Engineering and Department of Biological and Agricultural Engineering, Texas A$\&$M University, College Station, Texas 77843, USA}
\affiliation{Department of Physics and Astronomy, Texas A$\&$M University, College Station, Texas 77843, USA}
\affiliation{State Key Laboratory of Quantum Optics and Quantum Optics Devices, Collaborative Innovation Center of Extreme Optics, Shanxi University, Taiyuan 030006, China}

\author{Jinze Wu}
 \affiliation{Department of Physics, Zhejiang University, Hangzhou 310027, China}
 \affiliation{Institute for Quantum Science and Engineering and Department of Biological and Agricultural Engineering, Texas A$\&$M University, College Station, Texas 77843, USA}
 \affiliation{State Key Laboratory of Quantum Optics and Quantum Optics Devices, Collaborative Innovation Center of Extreme Optics, Shanxi University, Taiyuan 030006, China}
\author{Junxiang Zhang}
 \email[Author to whom all correspondence should be addressed: ]{junxiang\_zhang@zju.edu.cn}
 \affiliation{Department of Physics, Zhejiang University, Hangzhou 310027, China}
\author{Shiyao Zhu}
 \affiliation{Department of Physics, Zhejiang University, Hangzhou 310027, China}
\author{Girish S. Agarwal}
 \affiliation{Institute for Quantum Science and Engineering and Department of Biological and Agricultural Engineering, Texas A$\&$M University, College Station, Texas 77843, USA}
 \affiliation{Department of Physics and Astronomy, Texas A$\&$M University, College Station, Texas 77843, USA}

\date{\today}

\begin{abstract}
We show the existence of spin-Hall effect of light (SHEL) in an atomic medium which is made anisotropic via electromagnetically induced transparency. The medium is made birefringent by applying an additional linearly polarized coupling light beam. The refractive index and the orientation of the optics axis are controlled by the coupling beam. We show that after transmitting the atomic medium, a linearly polarized probe light beam splits into its two spin components by opposite transverse shifts. With proper choice of parameters and atomic density of about $2.5\times10^{17}\,\mathrm{m}^{-3}$, the shifts are about the order of wavelength and can be larger than the wavelength by increasing the atomic density. We propose a novel measurement scheme based on a balanced homodyne detection (BHD). By properly choosing the polarization, phase, and transverse mode of the local oscillator of the BHD,  one can independently measure (i) the SHEL shifts of the two spin components; (ii) the spatial and angular shifts; (iii) the transverse and longitudinal shifts. The measurement can reach the quantum limit of precision by detecting signals at the modulation frequency of the electro-optic modulator used to modulate the input probe beam. The precision is estimated to be at the nanometer level limited by the quantum noise. 
\end{abstract}

\pacs{42.50.Gy, 42.25.Lc}
\maketitle

%\tableofcontents

\section{Introduction}

Spin-orbit interaction of light \cite{Bliokh2015} originates from the fundamental spin properties of Maxwell’s equations, and thus is inherent in all basic optical processes. It yields many fascinating spin-dependent phenomena, in which the spin of light controls the propagation direction \cite{Bliokh2008, Bliokh2009, Bliokh2015_2, Gorodetski2012, Petersen2014, Junge2013, Mitsch2014}, the phase and intensity distribution \cite{Zhao2007, Marrucci2006, Marrucci2011, Bliokh2011, Sukhov2015}, etc. Spin-Hall effect of light (SHEL) \cite{Hosten2008, Onoda2004, Ling2017, Haefner2009, Aiello2009, Korger2014}, as a typical instance of the spin-orbit interaction of light, has been extensively investigated both theoretically and experimentally in a variety of systems, due to its potential applications in precision metrology \cite{Zhou2012, Zhou2012_1}, ultra-fast image processing \cite{Zhu2018}, etc. The SHEL manifests as the spin accumulation at the opposite sides of a light beam, or equivalently, the spin-dependent shifts of its centroid. Such shifts occur, for example,  when a light beam is reflected or refracted at a planar dielectric interface. Specifically, after reflection or refraction, light beams with right- and left-circular polarizations experience opposite shifts perpendicular to the incident plane, while a linearly polarized light beam splits into its two spin components. The former is the so-called Imbert-Fedorov (IF) shift \cite{Bliokh2013} and the latter is conventionally referred to as SHEL. Similar shifts also appear when a light beam propagates spirally in a smooth gradient-index medium \cite{Bliokh2008, Bliokh2009}. It has been proven that the total angular momentum, the spin angular momentum plus the orbit angular momentum, is conserved in the SHEL \cite{Bliokh2006}. The underlying physics of aforementioned SHEL is the spin-redirection Rytov-Vladimirskii-Berry (PVB) phase \cite{Bliokh2008_1} related to the change of the wave vector direction. Besides, another kind of geometric phase, i.e., the Pancharatnam-Berry (PB) phase \cite{Bliokh2008_1} associated with the manipulation of the polarization state, also leads to SHEL, where the light beam undergoes spin-dependent momentum shifts (deflection of the propagation direction) \cite{Bomzon2002, Shitrit2011}. 

The SHEL at an interface essentially arises from the interplay between the spin-redirection PVB phase and the sharply inhomogeneous refractive index. Actually, the anisotropic refractive index of, e.g. a uniaxial crystal, also induces the SHEL \cite{Bliokh2016}. This can be understood from the viewpoint of symmetry because both of them have cylindrically symmetric refractive indexes around the normal of the interface and the optics axis of the uniaxial crystal, respectively. However, in contrast to the interface whose symmetry is geometric, the symmetry of the uniaxial crystal is associated with the intrinsic anisotropy. Therefore, the incident angle is defined with respect to the optics axis. Note that this kind of anisotropy induced SHEL, which is also associated with the spin-redirection PVB phase, is different from the SHEL in a structured anisotropic medium, such as a space-variant subwavelength grating \cite{Bomzon2002}, which originates from the space-variant PB phase. Although the SHEL at an interface has been extensively studied, this anisotropy induced SHEL has received little attention.

In this article, we examine how an atomic medium can produce SHEL. This requires making the atomic medium anisotropic, which can be done by applying coupling laser fields and by using the electromagnetically induced transparency (EIT)  \cite{Fleischhauer2005, Patnaik2000}. Specifically, the atomic medium exhibits a linear birefringence when a linearly polarized coherent coupling beam is applied. When another linearly polarized probe beam passes through the atomic medium it experiences the SHEL shifts, which can be significantly enhanced by strong absorptive anisotropy. This is in close analogy to the giant transverse shifts near the Brewster angle at an interface \cite{Gotte2014, Luo2011, Xu2016}. We present detailed results not only for SHEL but other kind of shifts like Goos-H\"{a}nchen (GH) shifts and the angular shifts \cite{Bliokh2013}. Since the shifts are at the subwavelength scale, the experimental detection requires a high measurement precision. In previous investigations, the weak measurement \cite{Hosten2008, Aharonov1988, Duck1989, Ritchie1991, Dressel2014, Dennis2012, Zhou2012_2, Toppel2013, Goswami2014} is most widely used, which enlarges the shifts by a factor of several hundreds. Here, we provide a more efficient method to study the SHEL and related shifts based on a balanced homodyne detection (BHD) \cite{Hsu2004, Delaubert2006, Sun2014} with a properly chosen local oscillator (LO). We analyze the precision of this method and show that the minimum measurable shift is at the nanometer level under typical experimental conditions and is only limited by the quantum noise.

Our scheme shows three main advantages. First, differently from a natural crystal, e.g. calcite, the optical properties of the atomic medium can be flexibly controlled by the coupling beam. Especially, the orientation of the optics axis can be adjusted without moving the medium. In fact, in order to observe the SHEL solely induced by the anisotropy, the probe beam has to pass through the medium perpendicularly to the surface to avoid the influences of the interface induced SHEL, the GH shift, and the IF shift, and moreover, the optics axis has to be tilted. When the incident angle is varying, it is challenging to realize this in a natural crystal of which the optics axis is fixed, as the situation in Ref.~\cite{Bliokh2016}. Instead, this is straightforward to be implemented in our scheme. Second, the two spin components undergo a longitudinal shift and opposite transverse SHEL shifts, and all the shifts can be spatial and angular. That is to say, the shifts have three degrees of freedom: (i) the opposite shifts of the two spin components; (ii) the spatial and angular shifts; (iii) the longitudinal and transverse shifts. The traditional quadrant-detector-based measurement combined with the weak-value enhancement only detects the centroid shifts at the position of the detector, which is actually an overall effect of the shifts occurring in these three degrees of freedom. It is difficult for this scheme to fully decompose the overall shift in specific degrees of freedom. Especially, to decompose the spatial and angular shifts, one has to move the detector to perform the measurement at least in two positions. Since the shifts are very small, this probably decreases the precision. However, in our scheme, the shifts in the three degrees of freedom can be independently measured by appropriately choosing the polarization, phase, and transverse mode of the LO. Third, it has been proven that the quadrant detection is about 80\% efficiency compared to the BHD to measure the tiny shifts \cite{Delaubert2006, Sun2014}. In our scheme, the BHD-based measurement can reach the quantum limit by suppressing the classical noises using frequency modulation technique. 

\section{Anisotropic {EIT} medium}
\label{sec:2}

\begin{figure}[htb]
\centering
\includegraphics[width=\linewidth]{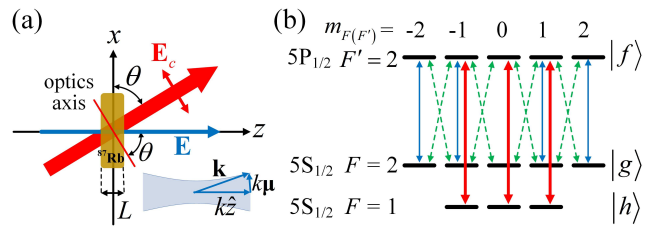}
\caption{
    (a) Schematic diagram of the system (see the text for the description). The lower right inset shows the distribution of the wave vector $\mathbf{k}=k(\bm{\mu}+\sqrt{1-|\bm{\mu}|^2}\hat{\bm{z}})$ narrowly around the central wave vector $k\hat{\bm{z}}$. (b) Relevant energy levels in the $^{87}$Rb D1 line. The thin solid blue and dashed green lines represent the $\pi$ and $\sigma^\pm$ transitions for the probe beam, respectively. The thick solid red lines represent the $\pi$ transitions for the linearly polarized coupling beam.
}
\label{fig1}
\end{figure}

We consider an atom-light interaction scheme, as shown in Fig.~\ref{fig1} (a), where the atomic medium with a length of $L$ is composed of $^{87}$Rb atoms whose D1 line consists of an excited state $\ket{f}=\ket{5\mathrm{P}_{1/2},F'=2}$, and two ground states $\ket{g}=\ket{5\mathrm{S}_{1/2},F=2}$ and $\ket{h}=\ket{5\mathrm{S}_{1/2},F=1}$, as illustrated in Fig.~\ref{fig1} (b). A monochromatic probe Gaussian beam propagating along the $z$-axis passes through the medium and acts on the transition $\ket{f}\leftrightarrow\ket{g}$. Its electric field $\mathbf{E}_p\propto\hat{\bm{\epsilon}}u(x,y)e^{-i\nu t+ikz}$, with $\hat{\bm{\epsilon}}$ the polarization vector, $\nu$ the frequency, $k=\nu/c$ the wave number, $c$ the speed of light in vacuum, $u(x,y)=w_0^{-1} e^{-(x^2+y^2)/w_0^2}$ the amplitude distribution in the position space, and $w_0$ the beam waist. Such a spatially confined light can be taken as a superposition of plane waves: $\mathbf{E}_p\propto\int \tilde{\mathbf{E}}_p\mathrm{d}k_x\mathrm{d}k_y=\hat{\bm{\epsilon}}\int \tilde{u}(k_x,k_y)e^{-i\nu t+i\mathbf{k}\cdot\mathbf{r}}\mathrm{d}k_x\mathrm{d}k_y$ with the position vector $\mathbf{r}=x\hat{\bm{x}}+y\hat{\bm{y}}+z\hat{\bm{z}}$ and the wave vector $\mathbf{k}=k\hat{\mathbf{k}}=k(\bm{\mu}+\sqrt{1-|\bm{\mu}|^2}\hat{\bm{z}})$ narrowly distributed around the central wave vector $k\hat{\bm{z}}$, where $\bm{\mu}=(k_x\hat{\bm{x}}+k_y\hat{\bm{y}})/k=\mu_x\hat{\bm{x}}+\mu_y\hat{\bm{y}}$, and $|\bm{\mu}|\ll1$, as shown in the inset of Fig.~\ref{fig1} (a). Here $\hat{\bm{x}}$, $\hat{\bm{y}}$, and $\hat{\bm{z}}$ are the unit vectors along the $x$-, $y$-, and $z$-axis, respectively. $\tilde{u}(k_x,k_y)=w_0/2e^{-w_0^2(k_x^2+k_y^2)/4}$ is the Fourier transform of $u(x,y)$, representing the amplitude distribution in the momentum space.

An additional linearly polarized strong coupling beam, resonant with the transition $\ket{f}\leftrightarrow\ket{h}$, is applied to prepare the atomic medium to be anisotropic. The coupling beam propagates along the direction at an angle $\theta$ to the $x$-axis, as illustrated in Fig.~\ref{fig1} (a). The beam width is large enough that in the medium the region where the atoms interact with the probe beam is fully covered by the coupling beam. We treat this coupling beam as a plane wave: $\mathbf{E}_c=\mathcal{E}_c\hat{\bm{\epsilon}}_ce^{-i\nu_c t+i\rm{k}_c\cdot\rm{r}}$ with $\mathcal{E}_c$ the amplitude, $\nu_c$ the frequency, $\hat{\bm{\epsilon}}_c=\hat{\bm{x}}\sin{\theta}-\hat{\bm{z}}\cos{\theta}$ the polarization vector, and $\bm{k}_c=k_c(\hat{\bm{x}}\cos{\theta}+\hat{\bm{z}}\sin{\theta})$ the wave vector.

Firstly, we consider the action of the atomic medium on the central plane-wave component of the probe beam and then show the SHEL in the next section. The quantization axis is taken to be parallel to $\hat{\bm{\epsilon}}_c$, and thus the coupling beam is $\pi$-polarized and couples the $\pi$ transitions between $\ket{f}$ and $\ket{h}$ as indicated by the red lines in Fig.~\ref{fig1} (b). For the special case of $\theta=\pi/2$, when the plane wave is linearly polarized along the $x$-axis (denoted by $\ket{x}$), i.e., parallel to the quantization axis, it couples the $\pi$ transitions between $\ket{f}$ and $\ket{g}$ as indicated by the blue lines. In this case, the system can be considered as a superposition of two three-level $\Lambda$-type EIT ($\ket{g,m_F=\pm1}\leftrightarrow\ket{f,m_{F'}=\pm1}\leftrightarrow\ket{h,m_F=\pm1}$) and two two-level ($\ket{g,m_F=\pm2}\leftrightarrow\ket{f,m_{F'}=\pm2}$) subsystems. The plane wave experiences a refractive index of $n_x=\sqrt{1+\chi_{\pi}}$ with the susceptibility
\begin{align}
\hspace*{-0.1cm}
    \chi_{\pi}=\sum_{i=\pm1}\kappa_{f_ig_i}\frac{\xi_{hg}\rho_{g_ig_i}}{\xi_{fg}\xi_{hg}-|\Omega_{f_ih_i}|^2} + \sum_{i=\pm2}\kappa_{f_ig_i}\frac{\rho_{g_ig_i}}{\xi_{fg}},
\label{eqs:susp_pi}
\end{align}
where the first (second) term denotes the susceptibilities of the three- (two-) level subsystems, and $\xi_{fg}=\delta-i\Gamma_{fg}$, $\xi_{hg}=\delta-i\Gamma_{hg}$, $\kappa_{f_ig_i}=N_a|\mathbf{d}_{f_ig_i}|^2/(\varepsilon_0\hbar)$, $\Omega_{f_ih_i}=\mathcal{E}_c\mathbf{d}_{f_ih_i}\cdot\hat{\bm{\epsilon}}_c/\hbar$ is the Rabi frequency of the coupling field, and $\rho_{g_ig_i}$ is the population of $\ket{g,m_F=i}$. Here $\delta=\omega_{fg}-\nu$ is the detuning of the probe beam with $\omega_{fg}$ the resonant frequency of transition $\ket{f}\leftrightarrow\ket{g}$, $\Gamma_{fg}=\gamma_{fg}$ and $\Gamma_{hg}\approx0$ are the decay rates of the off-diagonal density matrix elements $\rho_{fg}$ and $\rho_{hg}$, respectively, $2\gamma_{fg}$ is the spontaneous decay rate from $\ket{f}$ to $\ket{g}$,  $\mathbf{d}_{f_ig_i}$ and $\mathbf{d}_{f_ih_i}$ are the dipole matrix elements of the transitions $\ket{f,m_{F'}=i}\leftrightarrow\ket{g,m_F=i}$ and $\ket{f,m_{F'}=i}\leftrightarrow\ket{h,m_F=i}$, respectively, $N_a$ is the atomic number density, $\varepsilon_0$ is the vacuum permittivity, and $\hbar$ is reduced Planck constant.

When the plane wave is linearly polarized along the $y$-axis (denoted by $\ket{y}$), i.e., perpendicular to the quantization axis, it is a superposition of $\sigma^+$- and $\sigma^-$-polarized lights and thus couples the $\sigma^+$ and $\sigma^-$ transitions between $\ket{f}$ and $\ket{g}$ as indicated by the green lines in Fig.~\ref{fig1} (b). In this case, the system can be considered as a superposition of six three-level $\Lambda$-type EIT [$\ket{g,m_F=-2,-1,0}$ ($\ket{g,m_F=0,1,2}$)$\leftrightarrow\ket{f,m_{F'}=-1,0,1}\leftrightarrow\ket{h,m_F=-1,0,1}$] and two two-level ($\ket{g,m_F=\pm1}\leftrightarrow\ket{f,m_{F'}=\pm2}$) subsystems. The plane wave experiences a refractive index of $n_y=\sqrt{1+\chi_{\sigma}}$ with the susceptibility
\begin{align}
\hspace{-1cm}
    &\chi_{\sigma}=\frac{1}{2}\Bigg[\sum_{i=-2}^{0}\kappa_{f_{i+1}g_i}\frac{\xi_{hg}\rho_{g_ig_i}}{\xi_{fg}\xi_{hg}-|\Omega_{f_{i+1}h_{i+1}}|^2} + \kappa_{f_2g_1}\frac{\rho_{g_1g_1}}{\xi_{fg}}\nonumber\\
    &+\sum_{i=0}^{2}\kappa_{f_{i-1}g_i}\frac{\xi_{hg}\rho_{g_ig_i}}{\xi_{fg}\xi_{hg}-|\Omega_{f_{i-1}h_{i-1}}|^2} + \kappa_{f_{-2}g_{-1}}\frac{\rho_{g_{-1}g_{-1}}}{\xi_{fg}}\Bigg],
\label{eqs:susp_sigma}
\end{align}
where the first and third (second and forth) terms denote the susceptibilities of the three- (two-) level subsystems.

In the general case when $\theta$ is arbitrary, for the plane wave with polarization $\ket{x}$, its projections parallel and perpendicular to the quantization axis, which depend on $\theta$, couple the $\pi$ and $\sigma$ transitions between $\ket{f}$ and $\ket{g}$, respectively. However, for the plane wave with polarization $\ket{y}$, it is always perpendicular to the quantization axis regardless of $\theta$ and thus it still couples the $\sigma$ transitions. As a result, the refractive index for the $\ket{x}$ polarization is a function of $\theta$, while that for the $\ket{y}$ polarization is independent of $\theta$ (see Appendix \ref{App:A}):
\begin{align}
    n_x(\theta)=\sqrt{\frac{\varepsilon_{\sigma}\varepsilon_{\pi}}{\varepsilon_{\sigma}\sin^2{\theta}+\varepsilon_{\pi}\cos^2{\theta}}},\quad n_y=\sqrt{\varepsilon_{\sigma}},
\label{eqs:n}
\end{align}
where $\varepsilon_{\pi(\sigma)}=1+\chi_{\pi(\sigma)}$ is the relative permittivity. After passing through the atomic medium, the central plane waves with polarization $\ket{x}$ and $\ket{y}$ evolve as:
\begin{align}
    \ket{x}\rightarrow e^{i\phi_x(\theta)}\ket{x},\quad \ket{y}\rightarrow e^{i\phi_y}\ket{y},
\label{eqs:evol:cen}
\end{align}
where $\phi_x(\theta)=n_x(\theta)kL$ and $\phi_y=n_ykL$, of which the real and imaginary parts represent the phase shifts and the absorptions.

Without the coupling beam ($\Omega_{f_ih_i}=0$), $n_x=n_y$ and $n_x$ is independent of $\theta$, i.e., the atomic medium becomes isotropic. The presence of the coupling beam together with the difference of the transition strengths, which are determined by the dipole matrix elements, leads to $n_x(\theta)\ne n_y$ and the $\theta$-dependence, denoting that the refractive index depends on the polarization and the propagation direction of the probe beam. In other words, the atomic medium is linearly birefringent with an optics axis along the direction of the polarization of the coupling beam. It thus can be regarded as a controllable wave plate whose optics axis and refractive index are controlled by the coupling beam.

For the following results to be discussed, we have adopted experimentally feasible parameters: wavelength $\lambda=2\pi/k=794.98\,\mathrm{nm}$, reduced dipole matrix element $d=2.54\times10^{-29}\,\mathrm{C\cdot m}$, spontaneous decay rates $2\gamma_{fg}=2\gamma=2\pi\times5.75\,\mathrm{MHz}$, atomic number density $N_a=1.0\times10^{17}\,\mathrm{m}^{-3}$, atomic medium length $L=1\,\mathrm{mm}$, and reduced Rabi frequency of the coupling field $\Omega=\mathcal{E}_cd/\hbar=5\gamma$. The Rabi frequencies $\Omega_{f_ih_i}$ and the factors $\kappa_{f_ig_i}$ in Eqs.~(\ref{eqs:susp_pi}) and (\ref{eqs:susp_sigma}) can then be calculated using the Clebsch-Gordan coefficients (see Appendix \ref{App:B}). The population $\rho_{g_ig_i}$ is obtained by numerically solving the density-matrix equations using the AtomicDensityMatrix package provided in Ref.~\cite{ADM}.

\begin{figure}[htb]
\centering
\includegraphics[width=\linewidth]{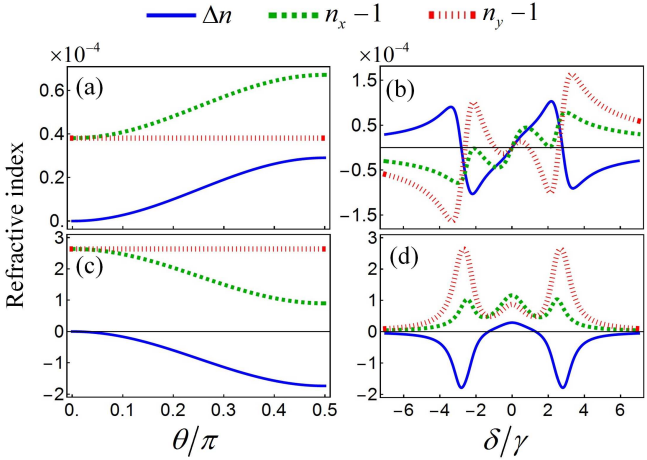}
\caption{
    (a), (b) Real and (c), (d) imaginary parts of $\Delta n$, $n_x-1$, and $n_y-1$ versus $\theta$ and $\delta$ with (a), (c) $\delta=2.7\gamma$ and (b), (d) $\theta=\pi/2$, respectively. See text for the details of the other parameters.
}
\label{fig2}
\end{figure}

\begin{figure*}[htb]
\centering
\includegraphics[width=\textwidth]{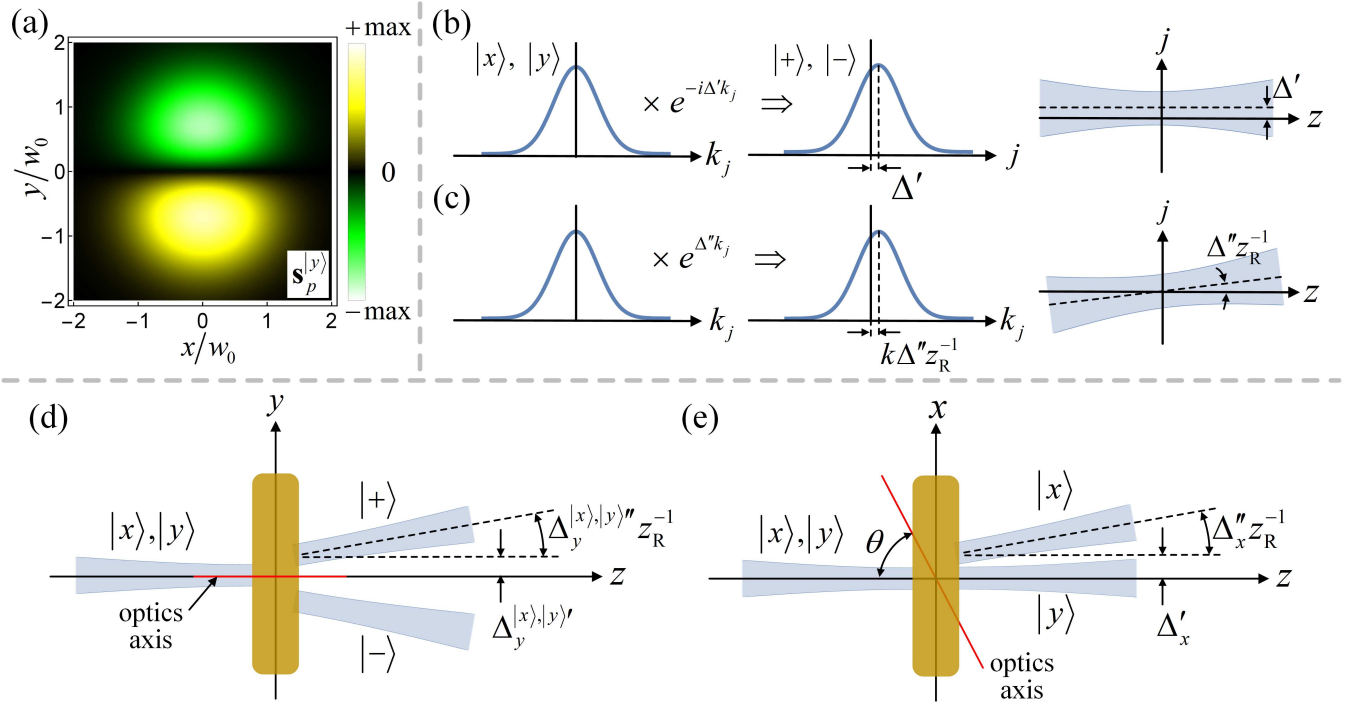}
\caption{
    Atomic SHEL and GH shifts. (a) Spin density distribution of $\mathbf{E}_p^{\ket{y}}$ showing the spin accumulation. (b) A phase modulation of $e^{-i\Delta'k_j}$ ($\Delta'\equiv\mathrm{Re}\Delta, j=x,y$) in the momentum space leads to a spatial shift of $\Delta'$ along the $j$-axis in the position space. (c) A amplitude modulation of $e^{\Delta''k_j}$ ($\Delta''\equiv\mathrm{Im}\Delta$) leads to a shift of $k\Delta''/z_{\mathrm{R}}$ along the $k_j$-axis in the momentum space which results in a angular shift of $\Delta''/z_{\mathrm{R}}$ in the $jz$-plane of the position space. (d) The transmitted probe beam splits into its two spin components $\ket{+}$ and $\ket{-}$ by opposite spatial shifts of $\pm\Delta_y^{\ket{x},\ket{y}'}$ and angular shifts of $\pm\Delta_y^{\ket{x},\ket{y}''}/z_{\mathrm{R}}$ along the $y$-axis. (e) The transmitted probe beam with $\ket{x}$ polarization experiences an overall spatial shift of $\Delta_x'$ and a angular shift of $\Delta_x''/z_{\mathrm{R}}$ along the $x$-axis, while the one with $\ket{y}$ polarization does not experience such shift. All the shifts are exaggerated.
}
\label{fig3}
\end{figure*}

The anisotropy of the atomic medium is characterized by the difference of $n_x(\theta)$ and $n_y$: $\Delta n(\theta)=n_x(\theta)-n_y$, of which the real and imaginary parts are responsible for the dispersive and absorptive anisotropies, respectively. When $\theta=0$, both the $\ket{x}$- and $\ket{y}$-polarization plane waves couple the $\sigma$ transitions between $\ket{f}$ and $\ket{g}$, and they experience the same refractive index, i.e., $n_x(0)=n_y=\sqrt{\varepsilon_{\sigma}}$ [see Eq.~(\ref{eqs:n})] and $\Delta n(0)=0$. When $\theta=\pi/2$, they couple the $\pi$ and $\sigma$ transitions, respectively, and one has $n_x(\pi/2)=\sqrt{\varepsilon_{\pi}}$, $n_y=\sqrt{\varepsilon_{\sigma}}$, and a maximum $\Delta n(\pi/2)$. $\Delta n(\theta)$ grows with $\theta$ ranging from 0 to $\pi/2$, as shown in Fig.~\ref{fig2} (a) and (c). From Fig.~\ref{fig2} (b) and (d), it is seen that the atomic medium exhibits weak anisotropy around $\delta=0$ and relative strong anisotropy near $\delta=\pm2.8\gamma$. The former (the latter) is due to the difference between the two- (three-) level-subsystem susceptibilities of $\chi_{\pi}$ and $\chi_{\sigma}$ [see Eqs.~(\ref{eqs:susp_pi}) and (\ref{eqs:susp_sigma})]. The shifts of $\pm2.8\gamma$ from the resonant frequency are a joint result of the Autler-Townes splitting \cite{Tannoudji1998} of $\pm\Omega_{f_{\pm1}h_{\pm1}}=\pm0.5\Omega=\pm2.5\gamma$ and $\pm\Omega_{f_0h_0}=\pm\Omega/\sqrt{3}\approx\pm2.9\gamma$ (see Appendix \ref{App:A}).

\section{Atomic SHEL and GH shifts}
\label{sec:3}

We now consider the action of the atomic medium on an arbitrary plane-wave component of the probe beam and show the SHEL. Equation (\ref{eqs:evol:cen}) needs slight modifications for $\bm{\mu}\ne0$. The plane-wave component with $\mu_x\ne0$ and $\mu_y=0$ propagates at a different angle $\theta+\mu_x$ to the optics axis, which slightly modifies the Eq. (\ref{eqs:evol:cen}): $\ket{x}\rightarrow e^{i\phi_x(\theta)}e^{-i\Delta_xk_x}\ket{x}$ and $\ket{y}\rightarrow e^{i\phi_y}\ket{y}$ with $\Delta_x=-k^{-1}\partial\phi_x(\theta)/\partial\theta$. If further $\mu_y\ne0$, the plane-wave component propagates in a slightly different plane, which can be obtained by rotating the $xz$-plane an angle $\mu_y/\sin{\theta}$ around the optics axis. Such a rotation induces spin-redirection PVB geometric phases of $\phi_{\mathrm{G}}=\mp\mu_y\cot{\theta}$ for the right- and left-circularly polarized plane-wave components, respectively, which is associated with the spin-orbit interaction \cite{Bliokh2015, Bliokh2013}. As a result, after the action of the atomic medium, $\ket{y}$ ($\ket{x}$) polarization is mixed into the plane-wave component with $\ket{x}$ ($\ket{y}$) polarization (see Appendix \ref{App:C}): $\ket{x}\rightarrow e^{i\phi_x(\theta)}e^{-i\Delta_xk_x}(\ket{x}+\Delta_y^{\ket{x}}k_y\ket{y})$ and $\ket{y}\rightarrow e^{i\phi_y}(\ket{y}-\Delta_y^{\ket{y}}k_y\ket{x})$ with $\Delta_y^{\ket{x}}=k^{-1}(1-e^{-i\phi(\theta)})\cot{\theta}$, $\Delta_y^{\ket{y}}=k^{-1}(1-e^{i\phi(\theta)})\cot{\theta}$, and $\phi(\theta)=\phi_x(\theta)-\phi_y$. Similar results for the interface can be found in Ref.~\cite{Hosten2008}.

\subsection{Spin Accumulation}
\label{sec:3:A}

In the momentum space, the probe beams with polarizations $\ket{x}$ and $\ket{y}$ evolve to
\begin{align}
    \tilde{\mathbf{E}}_p^{\ket{x}}&\propto e^{-i\Delta_xk_x}\tilde{u}\big(k_x,k_y\big)\big(\ket{x}+\Delta_y^{\ket{x}}k_y\ket{y}\big),\nonumber\\
    \tilde{\mathbf{E}}_p^{\ket{y}}&\propto\tilde{u}\big(k_x,k_y\big)\big(\ket{y}-\Delta_y^{\ket{y}}k_y\ket{x}\big),
\label{eqs:evol:lin:k}
\end{align}
respectively. The factors $e^{i\phi_x(\theta)}$ for $\tilde{\mathbf{E}}_p^{\ket{x}}$  and $e^{i\phi_y}$ for $\tilde{\mathbf{E}}_p^{\ket{y}}$ have been omitted for clarity. Transforming them back to the position space, we obtain (up to first order in $\Delta_x$, $\Delta_y^{\ket{x}}$ , and $\Delta_y^{\ket{y}}$)
\begin{align}
    \mathbf{E}_p^{\ket{x}}&\propto u\big(x-\Delta_x,y\big)\big(\ket{x}+2i\Delta_y^{\ket{x}}w_0^{-2}y\ket{y}\big),\nonumber\\
    \mathbf{E}_p^{\ket{y}}&\propto u\big(x,y\big)\big(\ket{y}-2i\Delta_y^{\ket{y}}w_0^{-2}y\ket{x}\big).
\label{eqs:evol:lin:r}
\end{align}
Due to the $y$-dependent terms, the transmitted probe beams are no longer linearly polarized. Instead, they exhibit a distribution of elliptical polarization. For example, when $\Delta_y^{\ket{y}}$  has a positive real part, the polarization of $\mathbf{E}_p^{\ket{y}}$  is left- (right-) handed elliptical polarization at $y>0$ ($y<0$). This effect can be well described by the spin density \cite{Bliokh2015_1,Aiello2015, Neugebauer2018}
\begin{align}
    \mathbf{s}_p^{\ket{x},\ket{y}}&=\frac{1}{4\nu}\mathrm{Im}\bigg[\varepsilon_0(\mathbf{E}_p^{\ket{x},\ket{y}})^*\times\mathbf{E}_p^{\ket{x},\ket{y}}\nonumber\\
    &+\mu_0(\mathbf{H}_p^{\ket{x},\ket{y}})^*\times\mathbf{H}_p^{\ket{x},\ket{y}}\bigg],
\label{eqs:s:r}
\end{align}
where $\mathbf{H}_p^{\ket{x},\ket{y}}=\sqrt{\varepsilon_0/\mu_0}\hat{\mathbf{k}}\times\mathbf{E}_p^{\ket{x},\ket{y}}$ is the magnetizing field and $\mu_0$ is the vacuum permeability. The spin densities of the linearly and circularly polarized fields are zero and maximum, respectively, while that of the elliptically polarized field is intermediate between them. Right- and left-circular (elliptical) polarizations have opposite spin densities. From Eqs. (\ref{eqs:evol:lin:r}) and (\ref{eqs:s:r}), we see that $\mathbf{s}_p^{\ket{x},\ket{y}}$ only has $z$-component:
\begin{align}
\hspace*{-0.1cm}
    \mathbf{s}_p^{\ket{x}}\propto\Delta_y^{\ket{x}'}yu^2(x-\Delta_x,y)\hat{\mathbf{z}},\,\,\,\mathbf{s}_p^{\ket{y}}\propto\Delta_y^{\ket{y}'}yu^2(x,y)\hat{\mathbf{z}},
\end{align}
with $\Delta_y^{\ket{x},\ket{y}'}\equiv\mathrm{Re}(\Delta_y^{\ket{x},\ket{y}})$, which implies a longitudinal spin. Fig.~\ref{fig3} (a) gives the spin density distribution $\mathbf{s}_p^{\ket{y}}$ of $\mathbf{E}_p^{\ket{y}}$ (the behavior of $\mathbf{s}_p^{\ket{x}}$ is almost the same as that of $\mathbf{s}_p^{\ket{y}}$). Obviously, the accumulation of the opposite spins occurs at the two sides of the transmitted probe beam along the $y$-axis, i.e., the SHEL is observed.

\subsection{SHEL and GH Shifts}
\label{sec:3:B}

The spin accumulation signifies opposite shifts along the $y$-axis of each spin components. Actually, in the spin basis $\ket{\pm}=(\ket{x}+i\ket{y})/\sqrt{2}$, Eq. (\ref{eqs:evol:lin:k}) can be rewritten as
\begin{align}
    \tilde{\mathbf{E}}_p^{\ket{x}}&\propto e^{-i\Delta_xk_x-i\Delta_y^{\ket{x}}k_y}\tilde{u}\big(k_x,k_y\big)\ket{+}\nonumber\\
    &+e^{-i\Delta_xk_x+i\Delta_y^{\ket{x}}k_y}\tilde{u}\big(k_x,k_y\big)\ket{-},\nonumber\\
    \tilde{\mathbf{E}}_p^{\ket{y}}&\propto e^{-i\Delta_y^{\ket{y}}k_y}\tilde{u}\big(k_x,k_y\big)\ket{+}\nonumber\\
    &-e^{i\Delta_y^{\ket{y}}k_y}\tilde{u}\big(k_x,k_y\big)\ket{-},
\label{eqs:evol:cir:k}
\end{align}
provided that $\big|\Delta_y^{\ket{x},\ket{y}}\big|\ll w_0$. In general, $\Delta_x$, $\Delta_y^{\ket{x}}$, and $\Delta_y^{\ket{y}}$  are complex. The real part $\Delta'\equiv\mathrm{Re}\Delta$ ($\Delta$ denotes $\Delta_x$, $\Delta_y^{\ket{x}}$, or $\Delta_y^{\ket{y}}$) corresponds to a phase modulation of $e^{-i\Delta'k_j}$ ($j=x,y$) in the momentum space, which leads to a spatial shift of $\Delta'$ along the $j$-axis in the position space: $\int e^{-i\Delta'k_j}\tilde{u}(k_j)\mathrm{d}k_j=u(j-\Delta')$, as illustrated in Fig.~\ref{fig3} (b). The imaginary part $\Delta''\equiv\mathrm{Im}\Delta$ corresponds to an amplitude modulation of $e^{\Delta''k_j}$, which results in a shift of $k\Delta''/z_{\mathrm{R}}$ along the $k_j$-axis in the momentum space: $e^{\Delta''k_j}\tilde{u}(k_j)\approx\tilde{u}(k_j-k\Delta''/z_{\mathrm{R}})$ with $z_{\mathrm{R}}=kw_0^2/2$ the Rayleigh range, or equivalently, an angular shift of $\Delta''/z_{\mathrm{R}}$ in the $jz$-plane of the position space: $\int e^{\Delta''k_j}\tilde{u}(k_j)\mathrm{d}k_j=u(j)e^{ikj\Delta''/z_{\mathrm{R}}}$, as shown in Fig.~\ref{fig3} (c). In the position space, we have
\begin{align}
    \mathbf{E}_p^{\ket{x}}&\propto u\big(x-\Delta_x,y-\Delta_y^{\ket{x}}\big)\ket{+}\nonumber\\
    &+u\big(x-\Delta_x,y+\Delta_y^{\ket{x}}\big)\ket{-},\nonumber\\
    \mathbf{E}_p^{\ket{y}}&\propto u\big(x,y-\Delta_y^{\ket{y}}\big)\ket{+}-u\big(x,y+\Delta_y^{\ket{y}}\big)\ket{-},
\label{eqs:evol:cir:r}
\end{align}
with the complex shifts of $\Delta_x$, $\Delta_y^{\ket{x}}$, and $\Delta_y^{\ket{y}}$. Equation (\ref{eqs:evol:cir:r}) indicates that the transmitted probe beam with $\ket{x}$ ($\ket{y}$) polarization indeed splits by opposite shifts of $\pm\Delta_y^{\ket{x}}$ ($\pm\Delta_y^{\ket{y}}$) along the $y$-axis into its two spin components $\ket{+}$ and $\ket{-}$, as shown in Fig.~\ref{fig3} (d). These shifts arise from the factors $e^{\pm i\Delta_y^{\ket{x}}k_y}$($e^{\pm i\Delta_y^{\ket{y}}k_y}$) which represent a coupling between the spin and the transverse momentum $k_y$. This spin-orbit interaction is a result of the geometric phase as mentioned above. $\Delta_x$ represents an overall longitudinal shift along the $x$-axis only for $\ket{x}$ polarization, as shown in Fig.~\ref{fig3} (e). It arises from the factor $e^{-i\Delta_xk_x}$ in Eq. (\ref{eqs:evol:cir:k}) which originates from the angular gradient of $n_x(\theta)$. This indicates that this shift is the atomic counterpart of the \textit{GH shift} of the reflected beam at an interface.

\begin{figure}[htb]
\centering
\includegraphics[width=\linewidth]{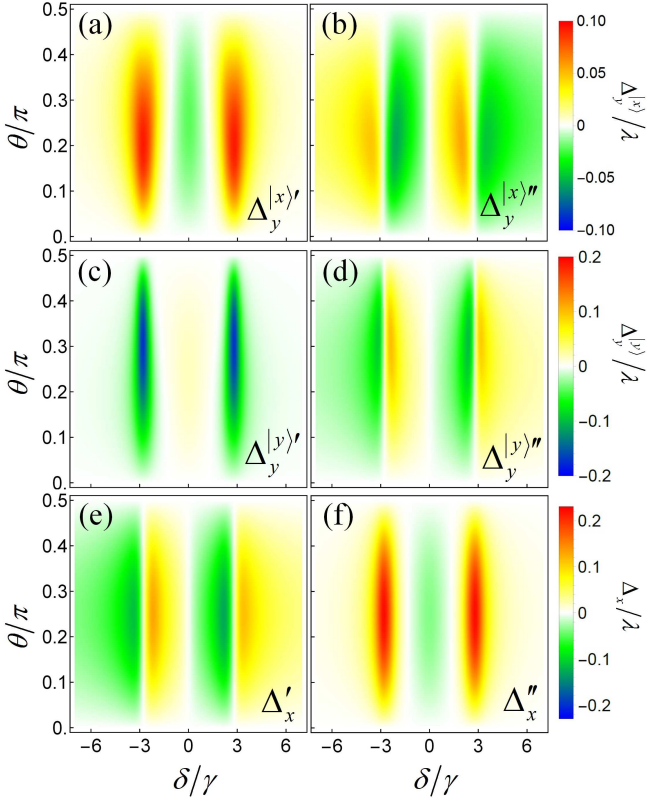}
\caption{
    (a)-(d) $\Delta_y^{\ket{x},\ket{y}}$ and (e)-(f) $\Delta_x$ versus $\delta$ and $\theta$. The other parameters are as in Fig.~\ref{fig2}. These are the quantities measured directly by the scheme shown in Fig.~\ref{fig6}
}
\label{fig4}
\end{figure}

\begin{figure}[htb]
\centering
\includegraphics[width=\linewidth]{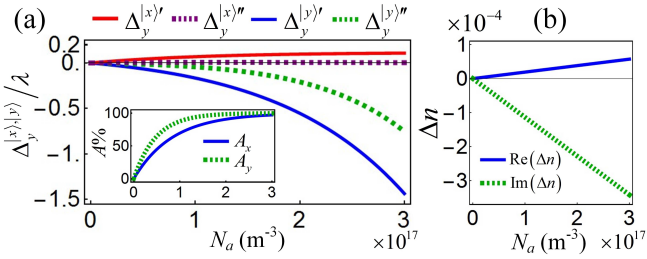}
\caption{
    (a) $\Delta_y^{\ket{x},\ket{y}}$, (inset) $A_{x,y}$, and (b) $\Delta n$ versus $N_a$ with $\delta=2.7\gamma$ and $\theta=0.3\pi$. The other parameters are as in Fig.~\ref{fig2}.
}
\label{fig5}
\end{figure}

Nonzero SHEL ($\Delta_y^{\ket{x},\ket{y}}\ne0$) requires two conditions: (i) the difference between the refractive indexes for the $\ket{x}$- and $\ket{y}$-polarization plane waves, i.e., $\Delta n(\theta)\ne0$; (ii) nonzero geometric phases for the spin components, i.e., $\phi_{\mathrm{G}}=\mp\mu_y\cot{\theta}\ne0$. For $\theta=0$ and $\pi/2$, one has $\Delta n=0$ and $\phi_{\mathrm{G}}=0$, respectively, and thus $\Delta_y^{\ket{x},\ket{y}}=0$, as shown in Fig.~\ref{fig4} (a)-(d). Nonzero $\Delta_x$ requires nonzero angular gradient of $n_x(\theta)$, i.e., $\partial n_x(\theta)/\partial\theta\ne0$. Similar to the SHEL, for $\theta=0$ and $\pi/2$, one has $\partial n_x(\theta)/\partial\theta=0$ [see Fig.~\ref{fig2} (a) and (c)] and thus $\Delta_x=0$, as shown in Fig.~\ref{fig4} (e)-(f).

Since these shifts are induced by the anisotropy, large shifts are expected around $\delta=0$ and $\pm2.8\gamma$, as shown in Fig.~\ref{fig4}. We note that the SHEL shifts increase remarkably with the absorptive anisotropy. In order to show this feature, here we give the explicit expressions of the SHEL shifts:
\begin{align}
    \Delta_y^{\ket{x}'}&=k^{-1}(1-e^{\phi''}\cos{\phi'})\cot{\theta},\nonumber\\
    \Delta_y^{\ket{x}''}&=k^{-1}e^{\phi''}\sin{\phi'}\cot{\theta},\nonumber\\
    \Delta_y^{\ket{y}'}&=k^{-1}(1-e^{-\phi''}\cos{\phi'})\cot{\theta},\nonumber\\
    \Delta_y^{\ket{y}''}&=-k^{-1}e^{-\phi''}\sin{\phi'}\cot{\theta},
\label{eqs:SHEL:shifts}
\end{align}
with $\phi'\equiv\mathrm{Re}\phi$ and $\phi''\equiv\mathrm{Im}\phi$, which represent the dispersive and absorptive anisotropies, respectively. The factor $e^{\phi''}$ ($e^{-\phi''}$) yields a significant enhancement of $\Delta_y^{\ket{x}}$ ($\Delta_y^{\ket{y}}$ ) for a large positive (negative) value of $\phi''$. The parameters used in the calculation yield a large negative value of $\phi''=\mathrm{Im}(\Delta n)kL$ near $\delta=\pm2.8\gamma$ [see Fig.~\ref{fig2} (d)], and thus a large $\Delta_y^{\ket{y}}$  [see Fig.~\ref{fig4} (c)-(d)]. A straightforward way to enhance $\phi''$ is to increase $N_a$, as shown in Fig.~\ref{fig5}, but the price to pay is a higher loss of the probe beam [see the inset of Fig.~\ref{fig5} (a) where the absorption is defined as $A_{x,y}=1-e^{-\mathrm{Im}\phi_{x,y}}$ for the $\ket{x}$- and $\ket{y}$-polarization, respectively]. As discussed in the next section, a lower transmitted power reduces the measurement precision. Therefore, there is a trade-off between the large SHEL shifts and the high precision.

\section{Measurement scheme of SHEL based on a BHD}

\begin{figure}[htb]
\centering
\includegraphics[width=\linewidth]{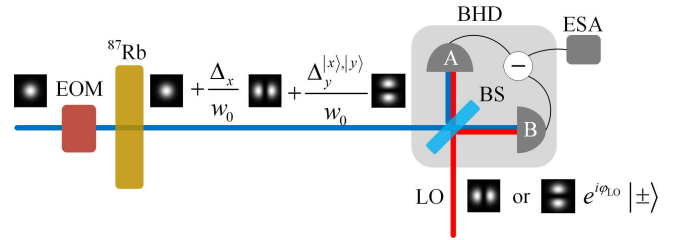}
\caption{
    Schematic diagram of the measurement scheme. The shifts of the transmitted probe beam are measured by the BHD with a TEM$_{10}$ (TEM$_{01}$) LO. The measurement is performed at RF frequency by using an EOM to modulate the probe beam.  EOM: electro-optic modulator; ESA: electronic spectrum analyzer; BS: 50/50 beam splitter.
}
\label{fig6}
\end{figure}

Obviously, the measurement of the tiny SHEL and GH shifts, at the subwavelength scale (see Fig.~\ref{fig4}), requires the sensitivity at the nanometer level. In addition, there are six shifts in all [Fig.~\ref{fig3} (d) and (e)], and thus a full characterization means independent measurement of each one. The traditional quadrant-detector-based measurement combined with the weak-value enhancement fails to meet the second requirement. Therefore, here we present a BHD-based measurement scheme, as schematically illustrated in Fig.~\ref{fig6}, which allows to independently measure each shift with a high precision.

Any spatial or angular shift of a fundamental Gaussian beam (TEM$_{00}$ mode) leads to excitations of higher Hermit-Gaussian modes (TEM$_{mn}$ modes). Up to first order, the electric field of the transmitted probe beam can be expanded as:
\begin{widetext}
\begin{align}
    \mathbf{E}_p^{\ket{x}}&\propto\sqrt{\frac{N_p}{2}}\left[u_{00}(x,y)+\frac{\Delta_x}{w_0}u_{10}(x,y)+\frac{\Delta_y^{\ket{x}}}{w_0}u_{01}(x,y)\right]\ket{+}\sqrt{\frac{N_p}{2}}\left[u_{00}(x,y)+\frac{\Delta_x}{w_0}u_{10}(x,y)-\frac{\Delta_y^{\ket{x}}}{w_0}u_{01}(x,y)\right]\ket{-},\nonumber\\
    \mathbf{E}_p^{\ket{y}}&\propto\sqrt{\frac{N_p}{2}}\left[u_{00}(x,y)+\frac{\Delta_y^{\ket{y}}}{w_0}u_{01}(x,y)\right]\ket{+}+\sqrt{\frac{N_p}{2}}\left[u_{00}(x,y)-\frac{\Delta_y^{\ket{y}}}{w_0}u_{01}(x,y)\right]\ket{-},
\label{eqs:HG}
\end{align}
\end{widetext}
where $u_{10}(x,y)$ and $u_{01}(x,y)$ denote, respectively, the TEM$_{10}$ and TEM$_{01}$ Hermit-Gaussian modes, and $N_p$ is the mean number of the photons detected in the time interval of one measurement, which is determined by the resolution bandwidth of the measurement device. Equation~(\ref{eqs:HG}) shows that the excited TEM$_{10}$ and TEM$_{01}$ modes carry the information of the shifts along the $x$- and $y$-axis, respectively. This can be extracted by using a TEM$_{10}$ or TEM$_{01}$ LO: $\mathbf{E}_{\mathrm{LO}}\propto\sqrt{N_{\mathrm{LO}}}u_{10,01}(x,y)e^{i\varphi_{\mathrm{LO}}}\ket{\pm}$ with $\varphi_{\mathrm{LO}}$ the phase, and $N_{\mathrm{LO}}$ the mean photon number. Since TEM$_{10}$ and TEM$_{01}$ modes are orthogonal to each other, a BHD with a TEM$_{10}$ (TEM$_{01}$) LO measures the shift along the $x$- ($y$-) axis. In addition, the polarization of the LO determines which spin component of the transmitted probe beam is detected. The transmitted probe beam and the LO are combined through a 50/50 beam splitter (BS). The two output beams of the BS are $\mathbf{E}_{\mathrm{A}}=(\mathbf{E}_p^{\ket{x},\ket{y}}+\mathbf{E}_{\mathrm{LO}})/\sqrt{2}$ and $\mathbf{E}_{\mathrm{B}}=(\mathbf{E}_p^{\ket{x},\ket{y}}-\mathbf{E}_{\mathrm{LO}})/\sqrt{2}$, respectively \cite{Bachor2004}. The mean number of the photons received by the detector A (B) is $N_{\mathrm{A(B)}}\propto\int \mathbf{E}_{\mathrm{A(B)}}^\dag\cdot\mathbf{E}_{\mathrm{A(B)}}\mathrm{d}x\mathrm{d}y$. The output of the BHD is given by the difference of the photocurrents of the two detectors: $I^{\mathrm{BHD}}\propto N_{\mathrm{A}}-N_{\mathrm{B}}$. For the TEM$_{01}$ and TEM$_{10}$ LO, we have
\begin{align}
    I_y^{\mathrm{BHD}}\propto&\pm\sqrt{2N_pN_{\mathrm{LO}}}\frac{1}{w_0}\big(\Delta_y^{\ket{x},\ket{y}'}\cos{\varphi_{\mathrm{LO}}}\nonumber\\
    &+\Delta_y^{\ket{x},\ket{y}''}\sin{\varphi_{\mathrm{LO}}}\big),\nonumber\\
    I_x^{\mathrm{BHD}}\propto&\sqrt{2N_pN_{\mathrm{LO}}}\frac{1}{w_0}\big(\Delta_x'\cos{\varphi_{\mathrm{LO}}}+\Delta_x''\sin{\varphi_{\mathrm{LO}}}\big).
\label{eqs:BHD}
\end{align}
It is clear that the spatial shifts $\Delta_y^{\ket{x},\ket{y}'}$ and $\Delta_x'$ (the angular shifts $\Delta_y^{\ket{x},\ket{y}''}$ and $\Delta_x''$) are measured when $\varphi_{\mathrm{LO}}=0$ ($\varphi_{\mathrm{LO}}=\pi/2$).

\begin{figure}[htb]
\centering
\includegraphics[width=0.75\linewidth]{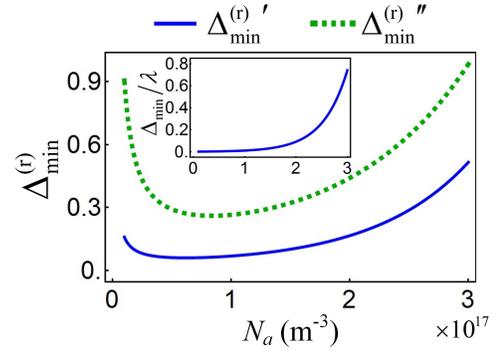}
\caption{
    $\Delta_{\mathrm{min}}^{(\mathrm{r})}$ and (inset) $\Delta_{\mathrm{min}}$ versus $N_a$ with $\delta=2.7\gamma$ and $\theta=0.3\pi$. See text for the parameters and the others are as in Fig.~\ref{fig2}.
}
\label{fig7}
\end{figure}

The tiny shifts to be measured are easily covered by various noises, such as technical noises in the laser source, mechanical and thermal noises in the optical setups, and vacuum noises. The classical noises can be effectively suppressed by performing the experiment at RF frequency (several MHz). Thanks to the strong dependences of the shifts on the frequency of the probe beam (see Fig.~\ref{fig4}), an RF-frequency modulation of the shifts can be easily achieved by modulating the frequency of the probe beam using, e.g. an electro-optic modulator (EOM). The output of the BHD should be measured at this modulation frequency using, e.g. an electronic spectrum analyzer (ESA). If the transmitted probe beam is assumed to be a coherent state, the signal-to-noise ratio (SNR) is given by $\mathrm{SNR}=2N_p\Delta^2/w_0^2$ \cite{Delaubert2006}. The minimum measurable spatial (angular) shift is estimated as the spatial (angular) shift with $\mathrm{SNR}=1$, yielding $\Delta_{\mathrm{min}}=w_0/\sqrt{2N_p}$ \cite{Delaubert2006}. Figure~\ref{fig7} shows, for example, the relative minimum measurable spatial (angular) shift $\Delta_{\mathrm{min}}^{(\mathrm{r})'}=\Delta_{\mathrm{min}}/|\Delta_y^{\ket{y}'}|$ ($\Delta_{\mathrm{min}}^{(\mathrm{r})''}=\Delta_{\mathrm{min}}/|\Delta_y^{\ket{y}''}|$) and $\Delta_{\mathrm{min}}$ (inset) as a function of $N_a$ with typical experimental parameters: the power of the incident probe beam $P_p=100\,\mu \mathrm{W}$; the beam waist of the probe beam $w_0=100\,\mu\mathrm{m}$; and the resolution bandwidth of the ESA $B_{\mathrm{res}}=100\,\mathrm{kHz}$. The number of the photons detected in the time interval $1/B_{\mathrm{res}}$ is $N_p=P_pe^{-2\mathrm{Im}\phi_y}/(B_{\mathrm{res}}\hbar\nu)$. It is clear that $\Delta_{\mathrm{min}}$ increases with $N_a$. However, as discussed in Sec.~\ref{sec:3}, $\Delta_y^{\ket{y}}$ is enhanced with a large $N_a$ [see Fig.~\ref{fig5} (a)]. The optimal $N_a$ is obtained when $\Delta_{\mathrm{min}}^{(\mathrm{r})'}$ and $\Delta_{\mathrm{min}}^{(\mathrm{r})''}$  are minimum. For the present parameters, the optimal $N_a$ is about $1\times10^{17}\,\mathrm{m}^{-3}$, and the minimum measurable spatial and angualr shifts angle are then estimated to be $\Delta_{\mathrm{min}}\approx9.0\,\mathrm{nm}$ and $\Delta_{\mathrm{min}}/z_{\mathrm{R}}\approx0.23\,\mu\mathrm{rad}$, respectively.

\section{Conclusions}

We have investigated the anisotropy induced SHEL with an anisotropic EIT atomic medium. We have shown that the medium exhibits a linear birefringence in the presence of a linearly polarized coherent coupling light beam, with the optics axis along the polarization direction of the coupling beam. After passing through the medium, a linearly polarized probe light beam experiences the spin accumulation and splits into its two spin components by opposite transverse shifts. Since the SHEL shifts are tiny (at the subwavelength scale), we have further presented a flexible measurement scheme based on a BHD. The scheme allows one to measure the shifts along any direction for each spin component by appropriately choosing the polarization and transverse mode of the LO. Besides, one can measure the spatial and angular shifts independently, when the phase of the LO is chosen to be 0 and $\pi/2$, respectively. The sensitive dependence of the SHEL shifts on the frequency of the probe beam makes it possible to perform the measurement at RF frequency such that the classical noises are efficiently suppressed and the precision can reach the quantum limit. Finally, we have shown that the minimum measurable shift is estimated to be at the nanometer level under typical experimental conditions.

\begin{acknowledgments}
We thank Dr. Jie Li for reading the paper and discussions. This work was supported by National Natural Science Foundation of China (91736209, 11634008, 11574188); Zhejiang Provincial Natural Science Foundation of China under Grant (No: LD18A040001). GSA thanks the Welch grant award No. A-1943-20180324 for support.
\end{acknowledgments}

\appendix

\section{Derivation of the refractive indexes}
\label{App:A}

Here we derive the refractive indexes [Eq. (3) in the main text] of the atomic medium for the probe beam. We start with Maxwell’s equations for a monochromatic wave \cite{Landau2013}:
\begin{align}
    \mathbf{B}=\frac{1}{c}\mathbf{n}\times\mathbf{E},\qquad\mathbf{D}=-\frac{1}{c\mu_0}\mathbf{n}\times\mathbf{B},
\label{eqs:Max}
\end{align}
with $\mathbf{E}$ the electric field, $\mathbf{D}$ the displacement field, $\mathbf{B}$ the magnetic field, and $\mathbf{n}=c\mathbf{k}/\nu$, where the magnitude of $\mathbf{n}$, denoted by $n$, represents the refractive index. Substituting the first equation into the second equation and using $\mathbf{D}=\varepsilon_0\bm{\varepsilon}\cdot\mathbf{E}$ and $c=1/\sqrt{\varepsilon_0\mu_0}$, we obtain
\begin{align}
    n^2\mathbf{E}-(\mathbf{n}\cdot\mathbf{E})\mathbf{n}-\bm{\varepsilon}\cdot\mathbf{E}=0.
\label{eqs:linsys}
\end{align}
The relative permittivity $\bm{\varepsilon}$ in the coordinate frame shown in Fig. 1 (a) in the main text is
\begin{align}
    \bm{\varepsilon}=\left(
        \begin{array}{ccc}
        \varepsilon_{\sigma}\cos^2{\theta}+\varepsilon_{\pi}\sin^2{\theta} & 0 & (\varepsilon_{\sigma}-\varepsilon_{\pi})\sin{\theta}\cos{\theta}\\
        0 & \varepsilon_{\sigma} & 0\\
        (\varepsilon_{\sigma}-\varepsilon_{\pi})\sin{\theta}\cos{\theta} & 0 & \varepsilon_{\sigma}\sin^2{\theta}+\varepsilon_{\pi}\cos^2{\theta}
        \end{array}
        \right).
\label{eqs:epsion}
\end{align}
Equation (\ref{eqs:linsys}) is a system of homogeneous linear equations. The compatibility condition is that the determinant of its coefficient matrix vanishes. Since the probe beam propagates along the $z$-axis, $\mathbf{n}=(0,0,n)$. Substituting Eq. (\ref{eqs:epsion}) into Eq. (\ref{eqs:linsys}) and setting the determinant of the coefficient matrix equal to zero, we obtain
\begin{align}
    (n^2-\varepsilon_{\sigma})(n^2\varepsilon_{\sigma}\sin^2{\theta}+n^2\varepsilon_{\pi}\cos^2{\theta}-\varepsilon_{\sigma}\varepsilon_{\pi})=0,
\label{eqs:Frensel}
\end{align}
and $n_x(\theta)$ and $n_y$ are two roots of $n$.

\section{Reduced dipole matrix elements and Rabi frequencies}
\label{App:B}

The dipole moment $\mathbf{d}=\mathbf{d}_1+\mathbf{d}_{-1}+\mathbf{d}_0$ has three components $\mathbf{d}_1=d_1\hat{\mathbf{u}}_1^*$, $\mathbf{d}_{-1}=d_{-1}\hat{\mathbf{u}}_{-1}^*$, and $\mathbf{d}_0=d_0\hat{\mathbf{u}}_0^*$ which couple the $\sigma^+$, $\sigma^-$, and $\pi$ transitions, respectively, where $\hat{\mathbf{u}}_{0,\pm1}$ are the spherical basis vectors with $\hat{\mathbf{u}}_0$ along the quantization axis. Using the Clebsch-Gordan coefficients and the Wigner-Eckart theorem, the dipole matrix element $\mathbf{d}_{f_{i+q}g_i(h_i)}=d_{f_{i+q}g_i(h_i)}\hat{\mathbf{u}}_q^*$ with $q=\pm1,0$, corresponding to the transition $\ket{f,m_{F'}=i+q}\leftrightarrow\ket{g(h),m_F=i}$, can be expressed as $d_{f_{i+q}g_i(h_i)}=M(F',F,m_F,q)d$ with $d$ the reduced dipole matrix element. The factor $M$ can be found in Ref.~\cite{Steck2015}. With these in hand, the factors $\kappa_{f_{i+q}g_i}=N_a|\mathbf{d}_{f_{i+q}g_i}|^2/(\varepsilon_0\hbar)$ in Eqs. (1) and (2) in the main text can be written as
\begin{align}
    \kappa_{f_{\pm2}g_{\pm2}}&=\frac{1}{3}\kappa,\,\,\,\,\kappa_{f_{\pm1}g_{\pm1}}=\frac{1}{12}\kappa,\nonumber\\
    \kappa_{f_{\pm1}g_{\pm2}}&=\kappa_{f_{\pm2}g_{\pm1}}=\frac{1}{6}\kappa,\,\,\,\,\kappa_{f_0g_{\pm1}}=\kappa_{f_{\pm1}g_0}=\frac{1}{4}\kappa,
\label{eqs:kappa}
\end{align}
with $\kappa=N_ad^2/(\varepsilon_0\hbar)$. The Rabi frequencies of the coupling field $\Omega_{f_ih_i}=\mathcal{E}_c\mathbf{d}_{f_ih_i}\cdot\hat{\bm{\epsilon}}_c/\hbar$ are
\begin{align}
    \Omega_{f_{\pm1}h_{\pm1}}=\frac{1}{2}\Omega,\qquad\Omega_{f_0h_0}=\frac{1}{\sqrt{3}}\Omega,
\label{eqs:Rabi}
\end{align}
with $\Omega=\mathcal{E}_cd/\hbar$ the reduced Rabi frequency of the coupling field.

\section{The action of the atomic medium on arbitrary plane-wave component}
\label{App:C}

\begin{figure}[htb]
\centering
\includegraphics[width=0.7\linewidth]{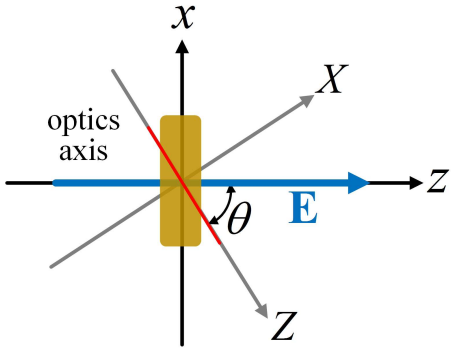}
\caption{
    The coordinate frames $xyz$ and $XyZ$.
}
\label{figS1}
\end{figure}

We introduce a new coordinate frame $XyZ$, which is obtained by rotating the coordinate frame $xyz$ an angle $-\theta$ around the $y$-aixs, as shown in Fig.~\ref{figS1}. The $Z$-axis is thus along the optics axis. For each plane-wave component, one can define a local coordinate frame attached to each $\mathbf{k}=k\hat{\mathbf{k}}$ with the unit vectors $\hat{\bm{z}}_{\mathbf{k}}=\hat{\mathbf{k}}$, $\hat{\bm{y}}_{\mathbf{k}}=\hat{\bm{Z}}\times\hat{\mathbf{k}}/|\hat{\bm{Z}}\times\hat{\mathbf{k}}|$, and $\hat{\bm{x}}_{\mathbf{k}}=\hat{\bm{y}}_{\mathbf{k}}\times\hat{\mathbf{k}}$, where $\hat{\bm{Z}}$ is the unit vector along the $Z$-axis. Geometrically, $\hat{\bm{x}}_{\mathbf{k}}$ ($\hat{\bm{y}}_{\mathbf{k}}$) is parallel (perpendicular) to the plane defined by $\hat{\mathbf{k}}$ and $\hat{\bm{Z}}$. With $\hat{\mathbf{k}}=\bm{\mu}+\sqrt{1-|\bm{\mu}|^2}\hat{\bm{z}}\approx\mu_x\hat{\bm{x}}+\mu_y\hat{\bm{y}}+\hat{\bm{z}}$ and $\hat{\bm{Z}}=-\sin{\theta}\hat{\bm{x}}+\cos{\theta}\hat{\bm{z}}$, one has $\hat{\bm{x}}_{\mathbf{k}}\approx\hat{\bm{x}}+\mu_y\cot{\theta}\hat{\bm{y}}-\mu_x\hat{\bm{z}}$ and $\hat{\bm{y}}_{\mathbf{k}}\approx-\mu_y\cot{\theta}\hat{\bm{x}}+\hat{\bm{y}}-\mu_y\hat{\bm{z}}$. The polarization of each plane-wave component is $\hat{\bm{\epsilon}}_{\mathbf{k}}=\hat{\bm{\epsilon}}-\hat{\mathbf{k}}(\hat{\mathbf{k}}\cdot\hat{\bm{\epsilon}})$ \cite{Merano2010}. For $\hat{\bm{\epsilon}}=\hat{\bm{x}}$ and $\hat{\bm{y}}$, we obtain
\begin{align}
    \hat{\bm{\epsilon}}_{\mathbf{k}}^{\hat{\bm{x}}}&\approx\hat{\bm{x}}-\mu_x\hat{\bm{z}}\approx\hat{\bm{x}}_{\mathbf{k}}-\mu_y\cot{\theta}\hat{\bm{y}}_{\mathbf{k}}\nonumber\\
    \hat{\bm{\epsilon}}_{\mathbf{k}}^{\hat{\bm{y}}}&\approx\hat{\bm{y}}-\mu_y\hat{\bm{z}}\approx\hat{\bm{y}}_{\mathbf{k}}+\mu_y\cot{\theta}\hat{\bm{x}}_{\mathbf{k}},
\end{align}
respectively. Under the action of the atomic medium, the plane-wave component with polarization $\hat{\bm{x}}_{\mathbf{k}}$ and $\hat{\bm{y}}_{\mathbf{k}}$ evolve as: $\hat{\bm{x}}_{\mathbf{k}}\rightarrow e^{i\phi_x(\theta+\mu_x)}\hat{\bm{x}}_{\mathbf{k}}\approx e^{i\phi_x(\theta)}e^{i\mu_x\partial\phi_x(\theta)/\partial\theta}\hat{\bm{x}}_{\mathbf{k}}$ and $\hat{\bm{y}}_{\mathbf{k}}\rightarrow e^{i\phi_y}\hat{\bm{y}}_{\mathbf{k}}$. Therefore, $\hat{\bm{\epsilon}}_{\mathbf{k}}^{\hat{\bm{x}}}$ and $\hat{\bm{\epsilon}}_{\mathbf{k}}^{\hat{\bm{y}}}$  evolve as:

\begin{align}
    \hat{\bm{\epsilon}}_{\mathbf{k}}^{\hat{\bm{x}}}&\rightarrow e^{i\phi_x(\theta)}e^{i\mu_x\partial\phi_x(\theta)/\partial\theta}\hat{\bm{x}}_{\mathbf{k}}-e^{i\phi_y}\mu_y\cot{\theta}\hat{\bm{y}}_{\mathbf{k}}\nonumber\\
    &\approx e^{i\phi_x(\theta)}e^{i\mu_x\partial\phi_x(\theta)/\partial\theta}\left[\hat{\bm{x}}+(1-e^{-i\phi(\theta)})\mu_y\cot{\theta}\hat{\bm{y}}\right],\nonumber\\
    \hat{\bm{\epsilon}}_{\mathbf{k}}^{\hat{\bm{y}}}&\rightarrow e^{i\phi_y}\hat{\bm{y}}_{\mathbf{k}}+e^{i\phi_x(\theta)}e^{i\mu_x\partial\phi_x(\theta)/\partial\theta}\mu_y\cot{\theta}\hat{\bm{x}}_{\mathbf{k}}\nonumber\\
    &\approx e^{i\phi_y}\left[\hat{\bm{y}}-(1-e^{i\phi(\theta)})\mu_y\cot{\theta}\hat{\bm{x}}\right],
\end{align}
where we have ignored the $z$-components. Using the notations in the main text, we obtain
\begin{align}
    \ket{x}&\rightarrow e^{i\phi_x(\theta)}e^{-i\Delta_xk_x}\left(\ket{x}+\Delta_y^{\ket{x}}k_y\ket{y}\right),\nonumber\\
    \ket{y}&\rightarrow e^{i\phi_y}\left(\ket{y}-\Delta_y^{\ket{y}}k_y\ket{x}\right),
\end{align}
with $\Delta_x$, $\Delta_y^{\ket{x}}$, $\Delta_y^{\ket{y}}$ defined in the main text.

% Bibliography
\bibliography{references}

\end{document}